\documentclass{article}
\usepackage{spconf,amsmath,graphicx}
\usepackage{lipsum}
\usepackage{comment}
\usepackage{url}
\usepackage{caption}
\usepackage[table,xcdraw]{xcolor}
\usepackage{booktabs}
\usepackage{adjustbox}
\usepackage{balance}
\usepackage{subfig}
\usepackage{multirow}
\usepackage[inline]{enumitem}
\usepackage{amsmath}

\title{On Crowdsourcing-Design with Comparison Category Rating for Evaluating Speech Enhancement algorithms}
\name{Angélica S. Z. Suárez $^1$, Clément Laroche $^{2}$, Line H. Clemmensen $^1$, Sneha Das $^1$ }
\address{
  $^1$ Dept. of Applied Mathematics and Computer Science, Technical University of Denmark, Denmark \\
  $^2$ GN Audio Research, Ballerup, Denmark\\
\\
claroche@jabra.com, sned@dtu.dk}

\begin{document}
\ninept
\maketitle
\begin{abstract}
%\lipsum[1]

Speech enhancement techniques improve the quality or the intelligibility of an audio signal by removing unwanted noise. It is used as preprocessing in numerous applications such as speech recognition, hearing aids, broadcasting and telephony. The evaluation of such algorithms often relies on reference-based objective metrics that are shown to correlate poorly with human perception. In order to evaluate audio quality as perceived by human observers it is thus fundamental to resort to subjective quality assessment and in doing so we identify subgroups of users where the subjective assessments correlate better to objective metrics.
In this paper, a user evaluation based on crowdsourcing (subjective) and the Comparison Category Rating (CCR) method is compared against the DNSMOS, ViSQOL and 3QUEST (objective) metrics.
The overall quality scores of three speech enhancement algorithms from real time communications (RTC) are used in the comparison using the P.808 toolkit.
Results indicate that while the CCR scale allows participants to identify differences between processed and unprocessed audio samples, two groups of preferences emerge: some users rate positively by focusing on noise suppression processing, while
others rate negatively by focusing mainly on speech quality.
We further present results on the parameters, size considerations and speaker variations that are critical and should be considered when designing the CCR-based crowdsourcing evaluation\footnote{The generated VoIP signals and the source codes are available at: \\\url{http://bit.ly/3lcFQqi}}.
\end{abstract}

\section{Introduction}
%\lipsum[1-5]
With the increasing use of hybrid work environments in recent years, there is a need for excellent audio quality in online communication, for both work and personal environments. Multimedia conferencing platforms, or Real-Time Communications (RTC), such as Teams, Zoom and Amazon Chime, are familiar Voice Over Internet Protocol (VoIP) systems used for these purposes. RTC commonly use noise suppressors to remove background noise, reverberation, and distortions from speech. However, they are also prone to adding artefacts and lowering the perceived quality of speech.
%\begin{comment}
\begin{figure}[!htb]
    \centering
    \includegraphics[width=0.89\columnwidth]{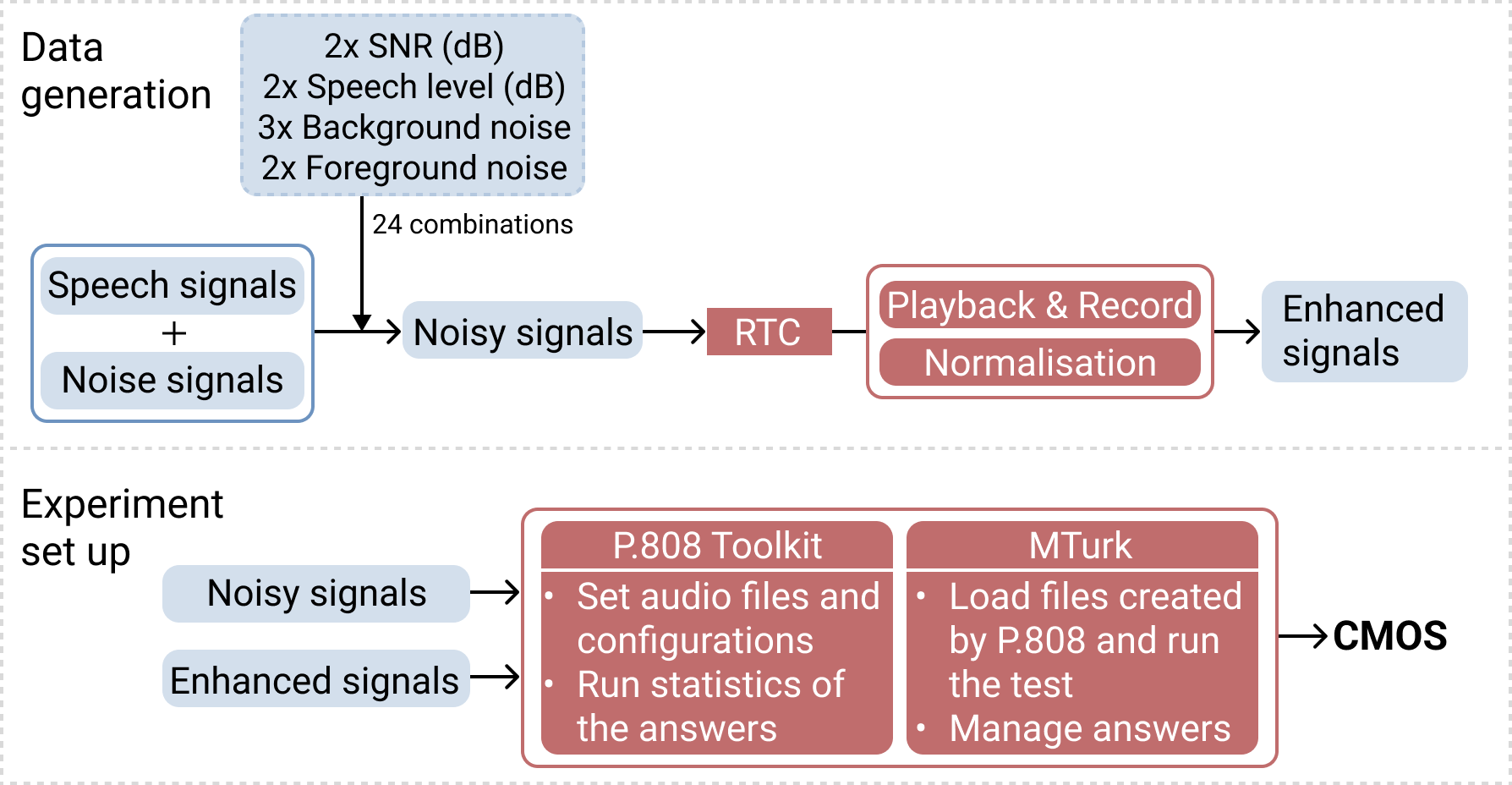}  % diagram_2_clear
    \caption{Pipeline of data generation and CCR experiment.}
    \label{fig:pipeline}

\end{figure}
%\end{comment}
Speech quality is commonly used by telecommunication service providers to optimize their enhancement algorithms. However predicting speech quality reliably is still an open problem~\cite{krishnamoorthy2011overview}. When the original clean signal is available, objective metrics such as perceptual evaluation of speech quality (PESQ) \cite{rix2001perceptual} and virtual speech quality objective listener (ViSQOL)~\cite{hines2012visqol}, have been widely used in the research community to benchmark speech enhancement algorithms \cite{duzinkiewicz2020overview}. These metrics are called full reference metrics or intrusive metric and they compare the degraded signal to the clean signal in a perceptual domain to obtain a measure based on a psycho-acoustic sound representation \cite{beerends1994perceptual,zilany2009phenomenological}. This measure of fit between the reference and the degraded is then mapped on a 1 to 5 scale using a cognitive model in order to predict the Mean Opinion Score (MOS). However, these objective metrics are shown to correlate poorly with human perception \cite{reddy2019scalable, paajanen2000new}. %Speech quality is used by telecommunication service providers in order to optimize the quality experienced by their customers.

% Therefore, challenges with respect to assessing speech quality reliability remain.

When the original clean signal is not available, assessing the quality of speech has to be carried out by means of subjective speech quality metrics, listening tests or crowdsourcing tests. Subjective metrics are normally designed to predict the MOS following the Telecommunication Sector of the International Telecommunication Union (ITU-T) Recommendation P.800 \cite{ITUP800}. There are three main test methods. Absolute Category Rating (ACR) consists of a single test condition (a snippet of audio) being presented to the users at a time. They should then give a quality rating on an ACR scale where Excellent equals to 5 and Bad equals to 1. The test conditions are presented in random order for each participant. The average score given by all participants, for each condition, is the MOS. In Degradation Category Rating (DCR) tests, the audience compares a processed sample~(B) with a reference sample~(A) on a five-point degradation category rating scale. Sound samples are always listened to in pairs of A–B or with repeating A–B–A–B. Finally, in Comparison Category Rating~(CCR) method, participants also listen to a reference and a processed stimuli but in random order~(A-B or B-A), and rate the quality of the second sample relative to the first sample on a scale ranging from -3 to 3~(a 7 points Likert scale). The sign of vote is later corrected to compare the quality of the processed sample to the unprocessed. The average of these votes leads to Comparison Mean Opinion Scores~(CMOS).

ACR, DCR and CCR procedures are traditionally conducted in a controlled environment to reduce the impact of variables unrelated to the purpose of the test. Crowdsourcing consists of carrying out quality assessment tasks in a crowd of Internet users, on a microtasking platform on which participants are paid for their service. The participants are given the freedom to choose the working environment and the listening devices which in turn increases the variability of the scores and provides less control over the quality of the answers \cite{naderi2021towards}. In contrast, subjective testing benefits greatly from crowdsourcing, including lower costs, higher speed, greater flexibility, scalability, and access to a diverse group of participants \cite{hossfeld2014best}.
ITU-T Rec. P.808 provides guidelines for the ACR paradigm on how to perform the speech quality assessment using crowdsourcing to reduce the variance of the test results due to the lack of a controlled environment. In addition, \cite{naderi2021speech} showed the reliability and validity of using CCR in the crowd for evaluating codec distortions.

%Listening tests are a known methodology to get subjective evaluations of audio quality \cite{loizou2011speech}. In these tests, human listeners score the quality of sound samples in a controlled environment, but due to the need for a large number of subjects to achieve statistically relevant results, they are very costly and time-consuming \cite{hu2007subjective}. Consequently, an ongoing research topic is the replacement of these tests with predictive models that can accurately estimate human evaluation regarding the quality of audio signals \cite{bech2007perceptual}.

%Finally, deep neural network based objective metrics trained with subjective human ratings collected using crowdsourcing, have gained popularity in recent years and appear as an alternative with a high correlation to subjective evaluation\cite{mittag2021nisqa, reddy2021dnsmos}.

The goal of this paper is to assess the suitability of the CCR test for evaluating speech enhancement algorithms. The contributions are: \begin{enumerate*}
\item We provide a database of pairs of clean, noisy and enhanced signals from three predominant VoIP solutions on the market.
\item We show that the user ratings are bimodal, whereby we recommend analysis models that account for the bimodality~(Sec.~3.1). To the best of our knowledge, this is the first work that presents empirical evidence on the bimodal phenomenon from experimental crowdsourced data.
\item We assess the suitability of the CCR methodology to evaluate speech enhancement algorithms and investigate CCR ratings against objective measures~(Sec.~3.2). Also, through our experimental design we identify subgroups of users where the subjective assessments correlate better to objective metrics.
\item We investigate the design choices in terms of the conditions, number of clips and speaker types and their influence on the crowd responses~(Sec.~3.3).
\end{enumerate*}

\section{Experimental Design}
\subsection{Dataset Design}
The overall data consists of clean speech signals, noise signals, noisy signals and enhanced signals of 10 seconds duration.
The clean speech signals created for the experiment are taken from the Voice Cloning ToolKit (VCTK) \cite{vctk2019voice}. The VCTK dataset consists of around 23 thousand utterances from 110 speakers recorded at 48kHz. From these clean speech files, 5 sentences from a male speaker and 5 sentences from a female speaker are selected.
Regarding the noises, Scaper\footnote{https://github.com/justinsalamon/scaper} \cite{salamon2017scaper} is used to create 8 noises of 10 seconds by mixing a continuous background noise (from office, cafeteria or park scenario) with transient noises common in these environments (see Table \ref{tab:foreground_noises}). The background noise data used is from the 2017 Detection and Classification of Acoustic Scenes and Events (DCASE) dataset \cite{mesaros2019sound}, while the transient noises are from the Freesound \cite{fonseca2017freesound} website.
The noisy signals are generated by adding the noise signals to the different clean speech files with a signal-to-noise ratio (SNR) of -3dB or 9dB. The speech level is previously adjusted to -30dB or -50dB SPL, which is a measure of the amplitude~($10\log10\sqrt{\frac{1}{N}\sum\limits_{i=1}^N x_i^2} = -50dB$ with signal values between -1 and 1). The total duration of transient noise in a clip is around 3s or 9s (30\% and 90\% of the total clip duration). There are 24 noisy conditions, defined by the multiple combinations of speech level, SNR, type of background noise and amount of transient noise targeted ($2\times2\times3\times2=24$); yielding 240 noisy clips ($24\times10$).

\begin{table}[!htb]
\small
\centering
\caption{Transient noise added over each background noise.}

\resizebox{0.99\columnwidth}{!}{
\begin{tabular}{@{}ccc@{}}
\toprule
Office &
  Cafeteria &
  Park \\ \midrule
\begin{tabular}[c]{@{}c@{}}Ringing, Cupboard-door\\ key \& Mouse clicks\end{tabular} &
  \begin{tabular}[c]{@{}c@{}}Stacking-cups\\ Microwave \& Slurping\end{tabular} &
  \begin{tabular}[c]{@{}c@{}}Birds, Footsteps\\ Car-horns\end{tabular} \\ \bottomrule
\end{tabular}
}

\label{tab:foreground_noises}
\end{table}

The enhanced signals are obtained by processing the noisy signals through the following RTCs: Teams\footnote{https://www.microsoft.com/en/microsoft-teams/group-chat-software}, Zoom\footnote{https://zoom.us/} and Amazon Chime\footnote{https://aws.amazon.com/es/chime/}, obtaining 720 enhanced signals in total and 72 test conditions (24 noisy conditions x 3 algorithms). The system comprised of two different computers to simulate a one-to-one online meeting, and a python script to playback and record the signals. The sender plays the noisy signals while the receiver records the signal processed by the RTC in real time.
The obtained enhanced signal depends on the characteristics of the input signal, the RTC used and the mode set for the noise suppression. A sampling rate of 48kHz and a noise suppression feature set to \textit{auto}, allow us to obtain wide-band audio for Teams, super wide-band audio for Zoom, and full-band audio for Chime.
%(using Teams and Zoom desktop applications and Chime web app).
We also found that Teams and Chime have an automatic gain control feature to always transmit the input signal at a fixed level. Therefore, the signals were normalised to have the same loudness prior to human rating. The visual modality was disabled for all tests.

\begin{comment}
Table X shows the output processing obtained for different input signals with the noise suppresion of the RTC set to \textit{auto} mode. The sampling rate for all the signals in the experiment is 48kHz.

\begin{table}[!htbp]
\small
\centering
\caption{Processing obtained per RTC over speech signals with different bandwidth (BW)}
\begin{tabular}{@{}cc|ccc@{}}
\toprule
\multicolumn{2}{c|}{Input} & \multicolumn{3}{c}{Output}   \\ \midrule
Sampling rate        & BW             & Teams  & Zoom     & Chime    \\ \midrule
8         & 4 (NB)         & 8 (WB) & 12 (SWB) & 8 (WD)   \\
16        & 8 (WB)         & 8 (WB) & 12 (SWB) & 12 (SWB) \\
48        & 24 (FB)        & 8 (WB) & 12 (SWB) & 20 (FB)  \\ \bottomrule
\end{tabular}
\end{table}
\end{comment}
\begin{figure}[!t]
  \centering
  \subfloat[]{\includegraphics[width=0.24\textwidth]{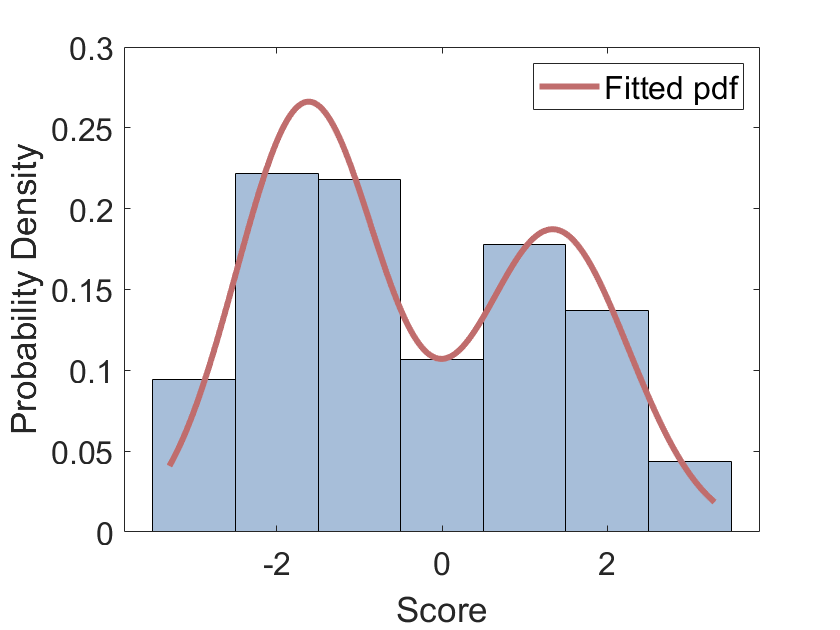}\label{fig:pos_neg_workers}}
  \hfill
 \subfloat[]{\includegraphics[width=0.24\textwidth]{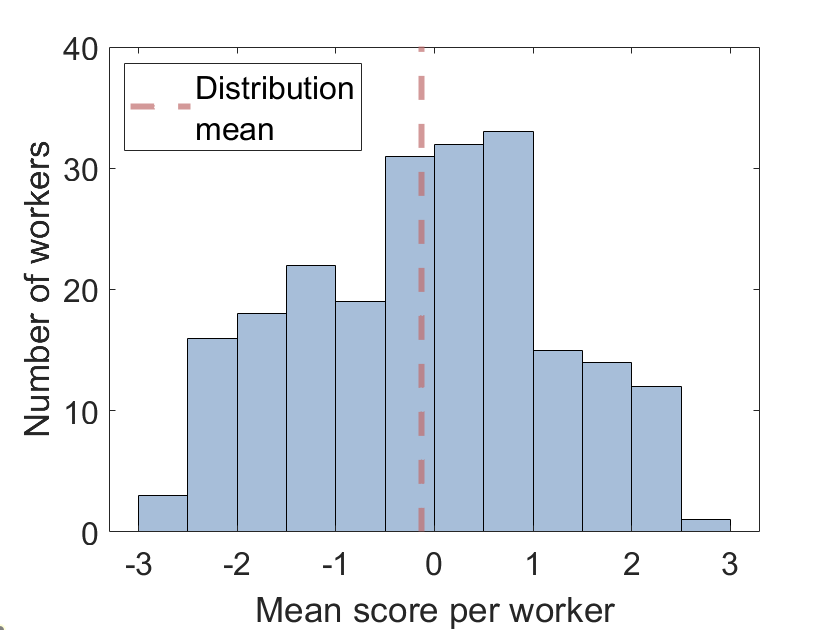}\label{fig:distribution}}
  \caption{(a) Probability density estimate of scores given by all workers and its distribution fitting. (b) Histogram of mean score per worker. %There are two different modes, representing a bimodal distribution. Workers whose mean scores are lower than this are labelled as “negative” and the rest as “positive”.
  }
\end{figure}
\subsection{Methods and tools}
% Explain experiments
We conducted subjective evaluation using crowdsourcing and CCR method. It was highlighted to users that overall quality included distortion or noise added/removed on/from the clip and their perceived speech quality. 8 pairs of clips were presented per assignment plus one gold question. The gold question presented the raters with the noisy signal as both, processed and unprocessed samples. The expected response was $0\pm1$~('sounded about the same'), and was used to filter usable responses.
% The listening order of the processed and unprocessed signals was random for each trial and participants did not know which sample was processed or unprocessed.

P.808 Toolkit \cite{naderi2020open} and Amazon Mechanical Turk\footnote{https://www.mturk.com/} (MTurk) were used to set up, publish and manage the CCR experiment and the answers collected.
% but also to manage the answers of the workers and get typical statistics from the scores.
% A csv file, with the links to the data, and an HTML file, containing the experiment interface, were generated by the P.808 Toolkit and loaded into Amazon MechanicalTurk (MTurk) prior to the release of the experiment.
Figure \ref{fig:pipeline} shows a diagram summarising the data generation step and the set up of the CCR experiment.
For the same dataset, objective metrics of different nature were also calculated: ViSQOL\footnote{https://github.com/google/visqol} was selected as an objective metric whose rating is mainly based on signal processing principles~\cite{hines2012visqol, hines2015visqol}; 3QUEST as a metric based on subjective listening test scores~\cite{processing2008transmission, duzinkiewicz2020overview}; and DNSMOS P.835\footnote{https://github.com/microsoft/DNS-Challenge} as a metric based on crowdsourcing ratings~\cite{reddy2021dnsmos, reddy2022dnsmos}. The scores of these objective metrics range from 1 to 5 (i.e. ACR scale).

\section{Results and Discussion}

%252 different workers participated in the CCR experiment and 731 assignments out of 810 were accepted to be used for the analysis after passing the data screening process. This step included a test for checking the use of both earpods, an environment suitability test, and a check on the variance score to be higher than 0.1. Overall, 5848 votes were collected (731 assignments x 8 clips), and on average there were 243 votes per condition.
% condition and algorithm: 81 votes

%With gold question activated:
216 workers participated in the CCR experiment and, 626 out of 810 assignments were accepted to be used for analysis after passing the data screening process. This step included a test of the correct use of both earpods, an environment suitability test and a gold standard question as specified in the ITU-T Rec. P.808 \cite{ITUP808}. Enough variance in the scores was also required. Overall, 5008 votes~(626 assignments$\times$8 clips) were collected and, on average, there were 69 votes per test condition (208 votes per noisy condition). To address the high variance in the crowd votes, we use CMOS per test condition computed as:

\begin{equation}
\label{eq:CMOS_condition}
    CMOS_{cond} = \frac{1}{M} \sum_{i=1}^{M} x_i \textrm{ with } x = \frac{\sum_{j=1}^{N} y_j}{N}
    % CMOS_{condition} = \frac{1}{MN} \sum_{j=1}^{M} \sum_{i=1}^{N} x_i
\end{equation}
where $M$ is the number of clips per condition, $x_i$ is the CMOS of a clip, $y_j$ is the vote given by a worker for a clip and $N$ is the number of votes per clip. The objective metrics DNSMOS, ViSQOL or 3QUEST are abbreviated in the analyses as DNS~(-S, B, O), ViS and 3Q~(-S, N, G), respectively~(speech, background/noise, overall).

\subsection{Bimodality in CMOS}
We begin by inspecting the distribution of scores of the crowd. The histogram of all the scores presented in Fig.~\ref{fig:pos_neg_workers} depicts a bimodal distribution, and the average score per worker in Fig.~\ref{fig:distribution} presents a division of preferences: some workers tend to rate positively while others rate negatively. We use linear fixed-effects models~(FEM) to study the influence of the modes on the correspondence to the objective measures in statistical terms, formulated as follows~\cite{chambers1992linear}:
\begin{equation}
\label{eq:all_posneg1}
Y_{ALL} = \alpha + \beta{C_{MOS}}+\epsilon,
\end{equation}
\begin{equation}
\label{eq:all_posneg2}
Y_{P-N} = \alpha + \beta{C_{MOS}}+\gamma F_{P-N}+\epsilon,
\end{equation}
\begin{table}[!htbp] \centering 
  \caption{Coefficients, (standard errors) and ANOVA between FEMs~(Eq.~\ref{eq:all_posneg1}), with positive-negative affiliation in Eq.~\ref{eq:all_posneg2}. The largest CMOS coefficients for each objective metric is emphasized.}
      \vspace{-0.15cm}
  \label{tab:allPosNeg} 
  \resizebox{0.89\columnwidth}{!}{
\begin{tabular}{@{\extracolsep{5pt}}lllll} 
\\[-1.8ex]\hline 
\hline \\[-1.8ex] 
 Obj-metric & Workers & \multicolumn{2}{c}{\textit{Factors~(Coeff~(std-error))}} & ANOVA \\ 
 \cline{1-5} 
 & & CMOS & Pos-Neg~$(=1)$ & Pr$(>F)$\\
  \cline{3-4} \\[-1.8ex] 
  \multirow{2}*{DNS-S} & All & .054~(.023)$^{**}$ & -- & --\\
  
  & P-N & {\bf.522~(.059)}$^{***}$ &$-$1.111~(.133)$^{***}$ & 4.9e-14$^{***}$\\
   \cline{2-5}  \\[-1.8ex] 
    \multirow{2}*{DNS-B} & All &  .050~(0.032) & -- & --\\
  & P-N & {\bf.491~(.094)}$^{***}$ &-1.045~(.211)$^{***}$ & 2.e-06$^{***}$\\
    \cline{2-5}  \\[-1.8ex] 
    \multirow{2}*{DNS-O} & All & .070~(.036)$^{**}$ & -- & --\\
  & P-N & {\bf.687~(.097)}$^{***}$&-1.461~(.218)$^{***}$ & 4.5e-10$^{***}$\\
      \hline \\[-1.8ex] 
      \multirow{2}*{ViS} & All & .057~(.024)$^{**}$& -- & --\\
  & P-N &  {\bf.51~(.062)}$^{***}$& -1.172~(.139)$^{***}$& 3.5e-14$^{***}$\\
        \hline \\[-1.8ex] 
      \multirow{2}*{3Q-S} & All & .059~(.033)$^{*}$ & -- & --\\
  & P-N &  {\bf.574~(.092)}$^{***}$& -1.22~(.207)$^{***}$& 2.65e-08$^{***}$\\
   \cline{2-5} 
      \multirow{2}*{3Q-N} & All & .052(.055) & -- & --\\
  & P-N &  {.50~(.16)}$^{***}$& -1.073~(.37)$^{***}$& 0.005$^{**}$\\
   \cline{2-5} 
        \multirow{2}*{3Q-G} & All &  .064$^{*}$(.035)& -- & --\\
  & P-N &  {\bf.63~(.098)}$^{***}$& -1.336~(.22)$^{***}$& 1.08e-08$^{***}$\\
  \hline 
\hline
\end{tabular} 
}
\end{table} 
where $\alpha$, $\beta$ and $\gamma$ are the coefficients of the fixed factors and $Y$, $C_{MOS}$ and $F_{P-N}$ are the objective metric, CMOS and positive/negative assignment of the CMOS, respectively. $\epsilon\sim N(0, \sigma^2)$ is the residual.
Further on, we perform analysis of variance~(ANOVA) between the fitted models to inspect if the model in Eq.~{\ref{eq:all_posneg2}} offers a significantly higher fit between the objective measures and the CMOS than the model in Eq.~{\ref{eq:all_posneg1}}~\cite{chambers1992linear}. The results illustrated in Tab.~\ref{tab:allPosNeg} depict that the fixed-effect model incorporating the positive-negative affiliation of the CMOS~(Eq.~\ref{eq:all_posneg2}): \begin{enumerate*}\item has larger and significant coefficients, and \item shows significantly better fit as per the p-value from the ANOVA model. Hence, including the positive-negative affiliation results in greater predictability of the objective measures.\end{enumerate*}

\begin{table}[!hbt]
\small
\addtolength{\tabcolsep}{-1pt}
\centering
\caption{Correlations between CMOS and MOS of objective metrics by average score per condition. Higher correlations of pos/neg workers are highlighted in \colorbox[HTML]{A3E6A3}{green}. The p-values of the correlations are lower than \( p < 0.05 \) except for CMOS-Noise case vs 3QUEST.}
\begin{adjustbox}{width=0.95\columnwidth}
\begin{tabular}{c|ccccccccc}
\hline
Metric & \multicolumn{9}{c}{CMOS}                                                                                                                                        \\ \hline
       & \multicolumn{3}{c|}{Speech}                               & \multicolumn{3}{c|}{Noise}                                & \multicolumn{3}{c}{Overall}             \\ \hline
       & \multicolumn{1}{c|}{All} & Neg & \multicolumn{1}{c|}{Pos} & \multicolumn{1}{c|}{All} & Neg & \multicolumn{1}{c|}{Pos} & \multicolumn{1}{c|}{All}  & Neg  & Pos  \\ \hline
DNS &
  \multicolumn{1}{c|}{0.69} &
  \cellcolor[HTML]{A3E6A3}0.68 &
  \multicolumn{1}{c|}{0.54} &
  \multicolumn{1}{c|}{0.47} &
  0.36 &
  \multicolumn{1}{c|}{\cellcolor[HTML]{A3E6A3}0.50} &
  \multicolumn{1}{c|}{0.57} &
  0.56 &
  0.50 \\
ViSQOL & \multicolumn{1}{c|}{--}  & --  & \multicolumn{1}{c|}{--}  & \multicolumn{1}{c|}{--}  & --  & \multicolumn{1}{c|}{--}  & \multicolumn{1}{c|}{0.67} & 0.65 & 0.57 \\
3QUEST &
  \multicolumn{1}{c|}{0.50} &
  \cellcolor[HTML]{A3E6A3}0.62 &
  \multicolumn{1}{c|}{0.31} &
  \multicolumn{1}{c|}{0.28} &
  {\color[HTML]{CB0000} 0.05} &
  \multicolumn{1}{c|}{\cellcolor[HTML]{A3E6A3}0.43} &
  \multicolumn{1}{c|}{0.53} &
  0.47 &
  0.48 \\ \hline
\end{tabular}
\end{adjustbox}
\label{tab:correlations}
\end{table}

A possible explanation for this phenomenon is that \textit{positive} workers like the noise suppression of the systems while \textit{negative} workers find the residual noise disturbing, or dislike the artefacts added to the enhanced signals \cite{neher2016investigating}. To assess this hypothesis, Spearman’s rank correlation and their corresponding p-values were computed over the mean scores per condition for all the scores, for only positive scores, and for only negative scores.
The results in Tab.~\ref{tab:correlations} show a positive correlation between the subjective scores and the different objective metrics. However, the correlations with the noise metrics are lower than with the speech metrics.
Also, taking into account the division of workers regarding subjective scores reveals that positive workers correlate better with noise metrics%(CMOS-DNS: \(0.50>>0.36\) and CMOS-3QUEST: \(0.43>>0.05\))
, confirming the statement that positive workers focus more on noise reduction than on speech quality. On the other hand, negative workers strongly correlate with speech metrics, meaning these workers focus mainly on the speech quality of the files. The single scores from ViSQOL correlate with both positive and negative workers.
\subsection{CCR-CMOS versus objective metric}
\begin{comment}
\begin{figure}[!htb]
    \centering
    \includegraphics[width=0.32\columnwidth]{images/DNSMOSI.pdf} \includegraphics[width=0.32\columnwidth]{images/visI.pdf}
\includegraphics[width=0.32\columnwidth]{images/3qI.pdf}
    \caption{Coefficient plot for CMOS~(Eq.~\ref{eq:FEM}). The p-values are shown by $^{*}p<0.05$; $^{**}p<0.0035$. Note that the x-axis ranges differ.}
    \label{fig:obj_vs_cmos}
    \vspace{-0.4cm}
\end{figure}
\end{comment}
\begin{figure}[!htb]
    \centering
    \includegraphics[width=0.98\columnwidth]{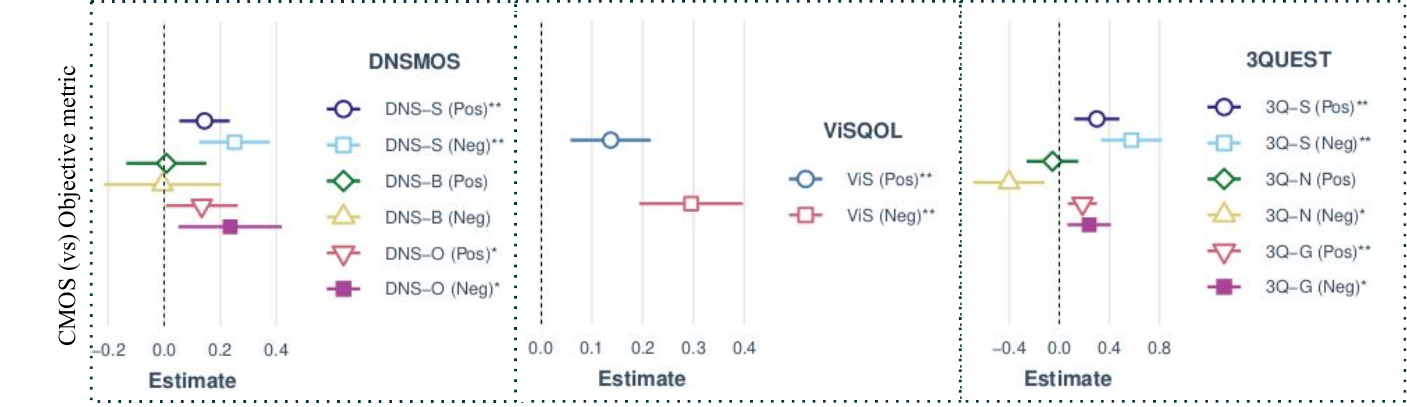}
    \caption{Coefficient plot for CMOS~(Eq.~\ref{eq:FEM}). The p-values are shown by $^{*}p<0.05$; $^{**}p<0.0035$. Note that the x-axes ranges differ.}
    \label{fig:obj_vs_cmos}
\end{figure}
The analysis models in the rest of the paper accounts for the positive-negative modalities of the scores.
To study the correspondence between the objective metric and the CMOS, we use linear FEM that incorporates all the condition parameters as follows:
\begin{equation}
\label{eq:FEM}
Y_{P-N} = \alpha + \beta{C_{MOS}}+\gamma{SNR}+\delta{S_L}+\lambda{N_E}+\eta{N_T}+\zeta{F_{alg}}+\epsilon,
\end{equation}
where $S_L$ is the speech level, $N_E$ is the amount of transient noise, $N_T$ is the type of background noise, $F_{alg}$ is the enhancement algorithm being evaluated and $\epsilon\sim N(0, \sigma^2)$ is the residual. $Y$ is the response variable representing the objective metrics. The coefficients and the standard errors for CMOS along with the p-values are illustrated in Fig.~\ref{fig:obj_vs_cmos}. With $\alpha=0.05$ and accounting for multiple testing, we apply Bonferroni correction whereby $\hat{\alpha}= \frac{\alpha}{14}=0.0035$. The adjusted-$R^2$ for all the comparisons are comparable. Our observations are that \begin{enumerate*}\item the CMOS has the highest, significant coefficient with respect to 3QUEST~(Neg) indicating that the CMOS corresponds best to 3QUEST~(Neg); \item In general, the scores from the negative mode have higher coefficients than the scores from the positive mode; \item CMOS from neither the positive nor the negative mode show significant predictability of DNS-B or 3Q-N. \end{enumerate*}
 %In contrast, noise type, amount of noise and speech level seems to influence specific objective metrics. The CMOS affiliated to the positive mode generally demonstrate higher coefficient than the negative CMOS mode.

\begin{comment}
\begin{figure*}[!htbp]
    \centering
        \subbottom{
          \includegraphics[width=0.9\linewidth]{images/SIG_BAK_plots.png}
          \label{fig:sig_bak}
        }
        ~
        \subbottom{
              \includegraphics[width=0.70\linewidth]{images/OVR_plots.png}
              \label{fig:ovr}
        }
        \caption{Correlations of the CMOS per condition and the MOS from DNSMOS, 3QUEST and ViSQOL metrics for positive and negative workers.}
    \label{fig:correlations_plots}
\end{figure*}
\end{comment}

\begin{comment}
\input{stat_anal_All_VS_posNegFact}
\end{comment}

\begin{comment}
\input{stat_anal_pos_neg}
\end{comment}

\subsection{Crowdsourcing-design based on statistical evidence}
\begin{table}[!t] \centering 
  \caption{Influence of the condition parameters on the CMOS responses as formulated in Eq.~\ref{eq:FEM_cond}.} 
  \vspace{-.25cm}
  \label{tab:effectofConditions} 
 \resizebox{0.89\columnwidth}{!}{
\begin{tabular}{@{\extracolsep{5pt}}lll} 
\\[-1.8ex]\hline 
\hline \\[-1.8ex] 
 Condition parameters & \multicolumn{2}{c}{$C_{MOS}$(Coefficient (Std-error))}\\ 
\cline{2-3} 
\\[-1.8ex] & Positive & Negative \\ 
\hline \\[-1.8ex] 
 SNR~(9dB) & {\bf 0.298~(0.079)}$^{****}$ & {\bf0.384~(0.054)}$^{****}$ \\ 
  Speech-level~(-30dB) & 0.099~(0.079) & {\bf0.260~(0.054)}$^{****}$ \\ 
  Noise-events~(90\%) & 0.016~(0.079) & 0.059~(0.054) \\ 
  Noise-type~(office) & 0.108~(0.096) & $-$0.110$^{*}$~(0.066) \\ 
  Noise-type~(park) & {\bf0.297~(0.096)}$^{****}$ & 0.014~(0.066) \\ 
  Algorithm~(2) & $-$0.243~(0.096)$^{***}$ & 0.134~(0.066)$^{**}$ \\ 
  Algorithm~(3) & $-$0.023~(0.096) & 0.041~(0.066) \\ 
  %Constant & 0.492~(0.111)$^{****}$ & $-$1.760~(0.076)$^{****}$ \\ 
 
 \hline \\[-1.8ex] 
%Observations & 72 & 72 \\ 
Adjusted R$^{2}$ & 0.271 & {\bf0.520} \\ 
\hline 
\hline \\[-1.8ex] 
\textit{Note:}  & \multicolumn{2}{r}{$^{*}$p$<$0.1; $^{**}$p$<$0.05; $^{***}$p$<$0.025; $^{****}$p$<$0.01} \\ 
\end{tabular}
}
\vspace{-0.5cm}
\end{table} 
In the following sections, we will present results on \begin{enumerate*}\item condition parameters that are significant in the study, \item speakers' influence on user ratings and objective measures, \item impact of number of clips on the predictability of CMOS. \end{enumerate*}\\
\noindent
{\bf Influence of condition parameters:}
We employ FEM to study the relationship between the obtained CMOS from the crowd and the condition parameters, whereby we incorporate the parameters in each condition as a factor in the model:
\begin{equation}
\label{eq:FEM_cond}
C_{MOS_{P-N}} = \alpha + \beta{SNR}+\gamma{S_L}+\delta{N_E}+\eta{N_T}+\zeta{F_{alg}}+\epsilon.
\end{equation}
We compute the models separately for the positive and negative modes. With $\alpha=0.05$ and accounting for multiple testing, we apply Bonferroni correction and obtain $\hat{\alpha}= \frac{\alpha}{2}=0.025$. The coefficients and standard-errors of each condition is shown in Tab.~\ref{tab:effectofConditions} in addition to the p-values. We observe the following: \begin{enumerate*}\item SNR shows large coefficients for both modes, hence indicating higher influence on the crowd responses. \item Speech-level seems to effect the negative mode, and \item noise-type, specifically park shows higher predictability for the positive mode. \item From the adjusted-$R^2$ values, the condition parameters account for lower variance in the positive mode than in the negative mode, hence indicating the different preferences of the crowd in each mode.\end{enumerate*}\\
\noindent
{\bf Influence of speakers on responses and objective metric:}
Two speakers~(one male, one female) were used for the 10 speech clips.
To study the influence of speakers on both, the CMOS responses and the objective measures, we modify the formulation in Eq.~\ref{eq:FEM} to incorporate the clips as a random effect~($\theta F_{Cl}\sim N(0, \sigma_{Cl})$) and the speaker/gender~($F_{Sp}$) as a fixed effect as follows:
\begin{equation}\label{eq:MEM_gen}
\begin{split}
Y &= \alpha + \beta{C_{MOS}}+\gamma{SNR}+\delta{S_L}+\zeta{N_E}+\eta{N_T}\\
&\qquad+ \lambda F_{P-N}+\nu F_{Sp}+\theta F_{Cl}+\epsilon
\end{split}
\end{equation}
thereby employing a linear mixed-effects model~(LMER)~\cite{pinheiro2006mixed}. We observed {\it no} statistical differences in the CMOS responses based on the speaker. However, the observations presented in Tab.~\ref{tab:sta_clip_level} indicate that the speaker seems to have a significant effect on DNSMOS. Note that with $\alpha=0.05$ and accounting for multiple testing via Bonferroni correction the corrected value is $\hat{\alpha}= \frac{\alpha}{7}=0.0071$. Since only two speakers~(one male and female each) were used in the study, this needs to be further investigated with larger variation in speakers.
\begin{table}[!htb] \centering 
  \caption{Coefficients, standard errors for random-effects model with speaker factor Eq.~{\ref{eq:MEM_gen}}.}
  \vspace{-.25cm}
  \label{tab:sta_clip_level} 
  \resizebox{0.89\columnwidth}{!}{
\begin{tabular}{@{\extracolsep{5pt}}llll} 
\\[-1.8ex]\hline 
\hline \\[-1.8ex] 
 Obj-metric & \multicolumn{3}{c}{\textit{Factors~(Coefficient (Std-error))}} \\ 
 \cline{1-4} 
 & CMOS & Pos-Neg~$(=1)$ & Speaker/Gender(M)\\
  \cline{2-4} \\[-1.8ex] 

  DNS-S & .038~(.008)$^{***}$ &.08~(.02)$^{***}$ & {\bf.38~(.09)}$^{***}$\\
   %\cline{2-4}  \\[-1.8ex] 
  DNS-B & .013~(.012) &-.027~(.03) & {\bf.21~(.05)}$^{***}$\\
    %\cline{2-4}  \\[-1.8ex] 
 DNS-O & .04~(.01)$^{***}$ &-.08~(.03)$^{**}$ & {\bf.36~(.10)}$^{**}$\\
      \hline \\[-1.8ex] 
  ViS &  .03~(.007)$^{***}$& -.07~(.018)$^{***}$& .04~(.04)\\
        \hline \\[-1.8ex] 
  3Q-S &  .08~(.017)$^{***}$&-.17~(.05)$^{***}$& -0.17~(.10)\\
   %\cline{2-4} 
 3Q-N &  0.02~(.02)& -.05~(.06)& .11~(.04)$^{**}$\\
   %\cline{2-4} 
 3Q-G&  .06~(.02)$^{***}$& -.14~(.04)$^{***}$& -.06~(.08)\\
  \hline 
\hline\\[-1.8ex] 
\textit{Note:}  & \multicolumn{3}{l}{$^{**}$p$<$0.05; $^{***}$p$<$0.0071} \\ 
\end{tabular} 
}
\vspace{-.15cm}
\end{table} 

\begin{figure}[!htb]
    \centering
    \includegraphics[width=0.45\columnwidth]{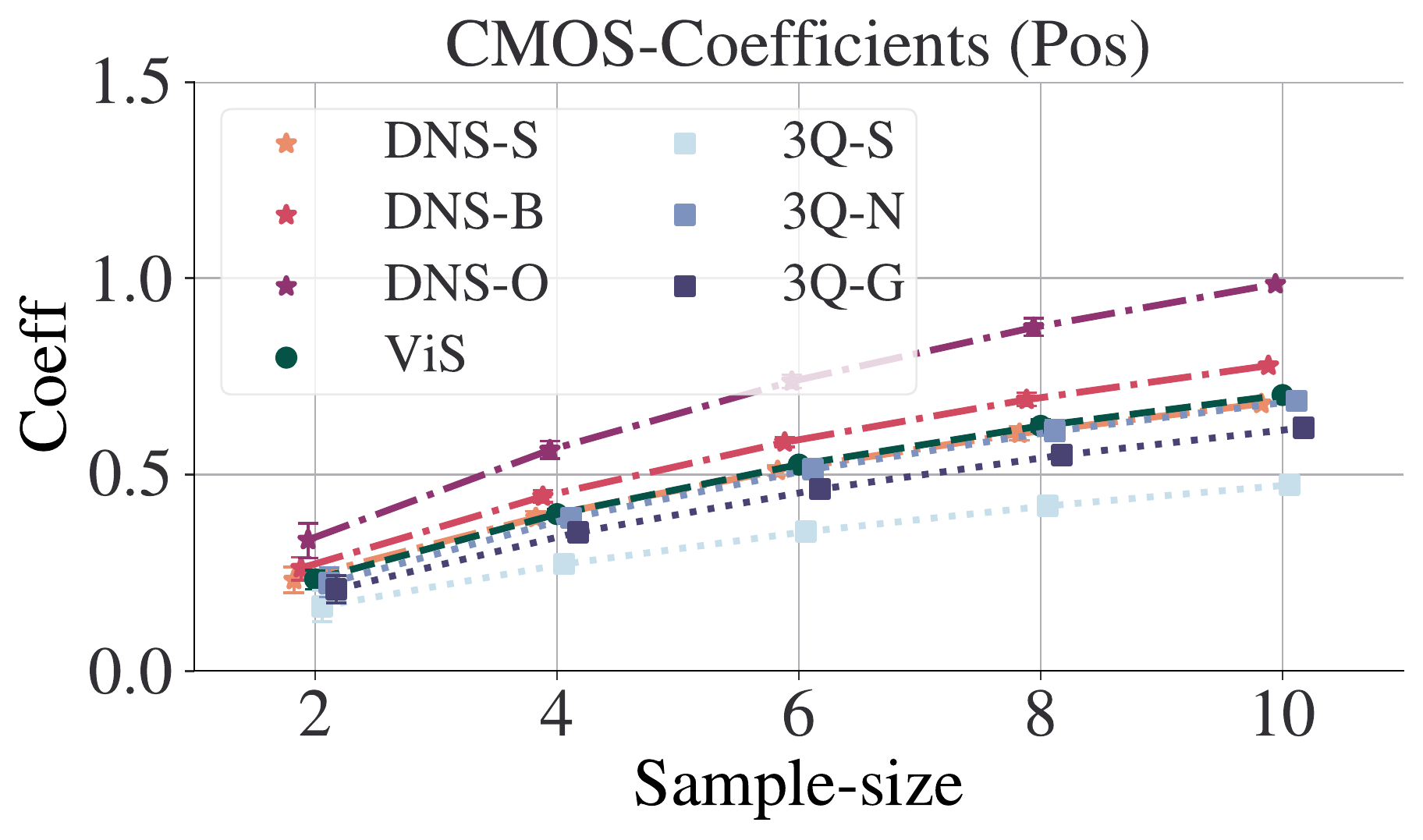} \includegraphics[width=0.45\columnwidth]{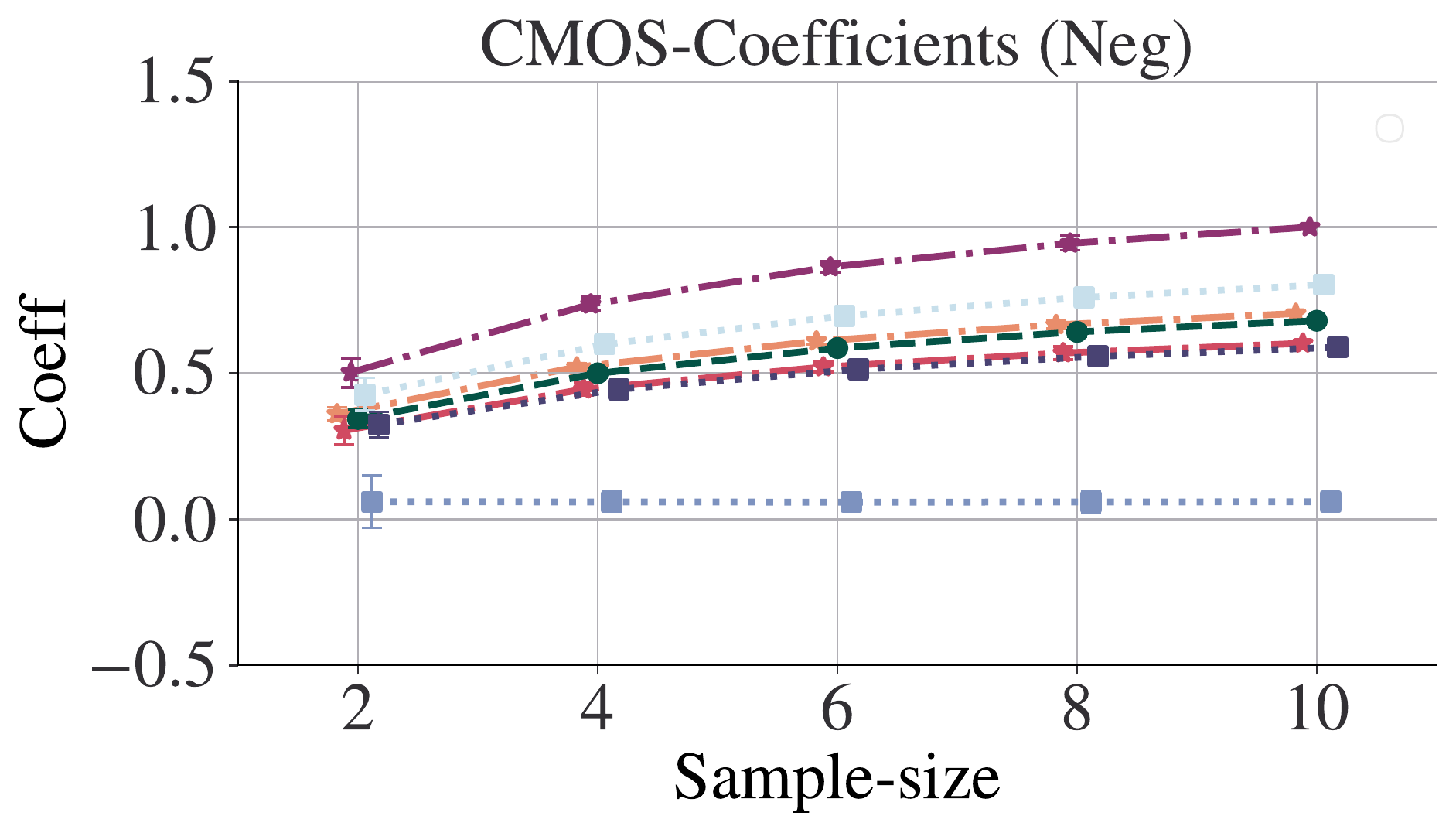} \\% diagram_2_clear|
        \includegraphics[width=0.45\columnwidth]{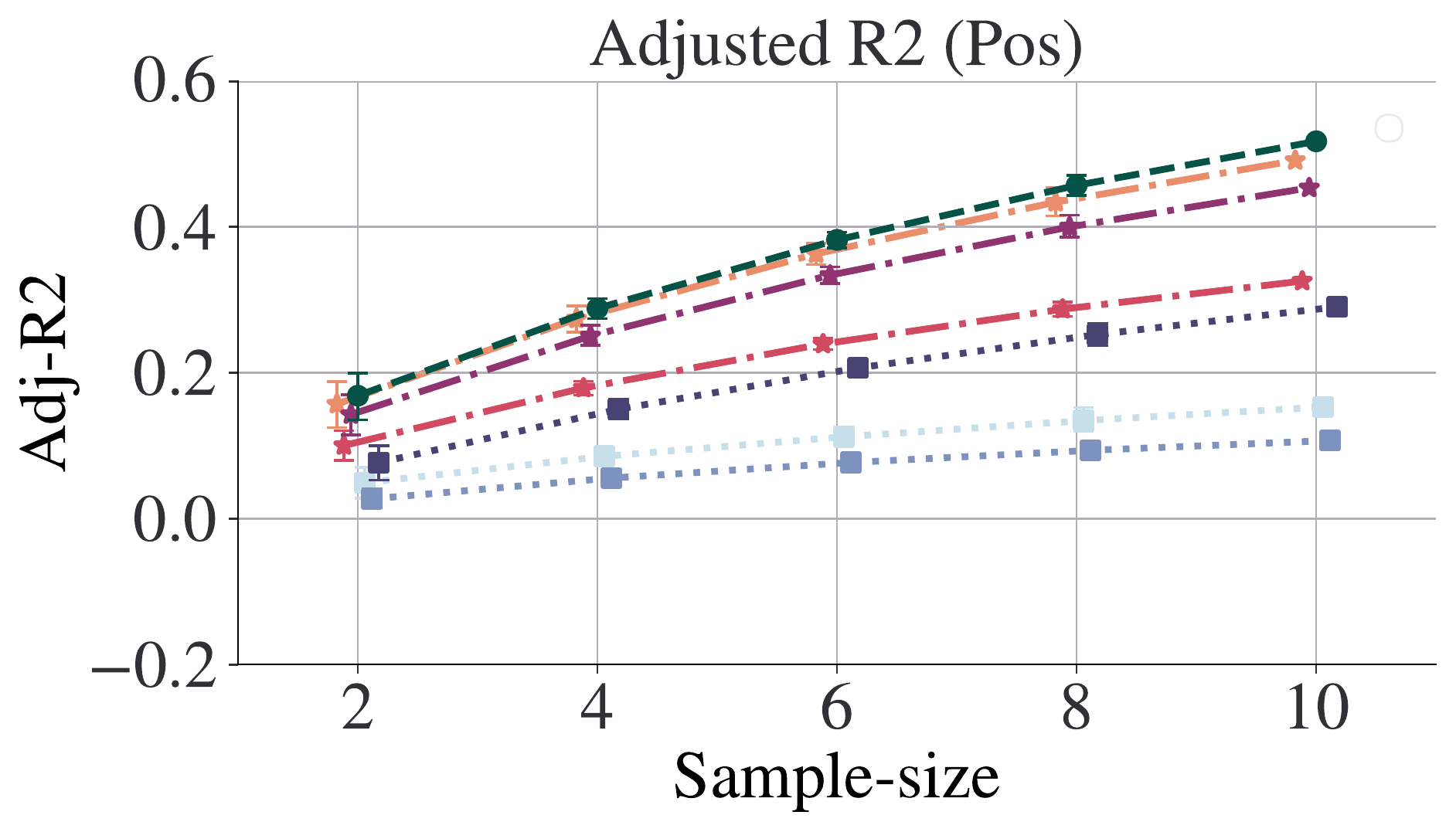} \includegraphics[width=0.45\columnwidth]{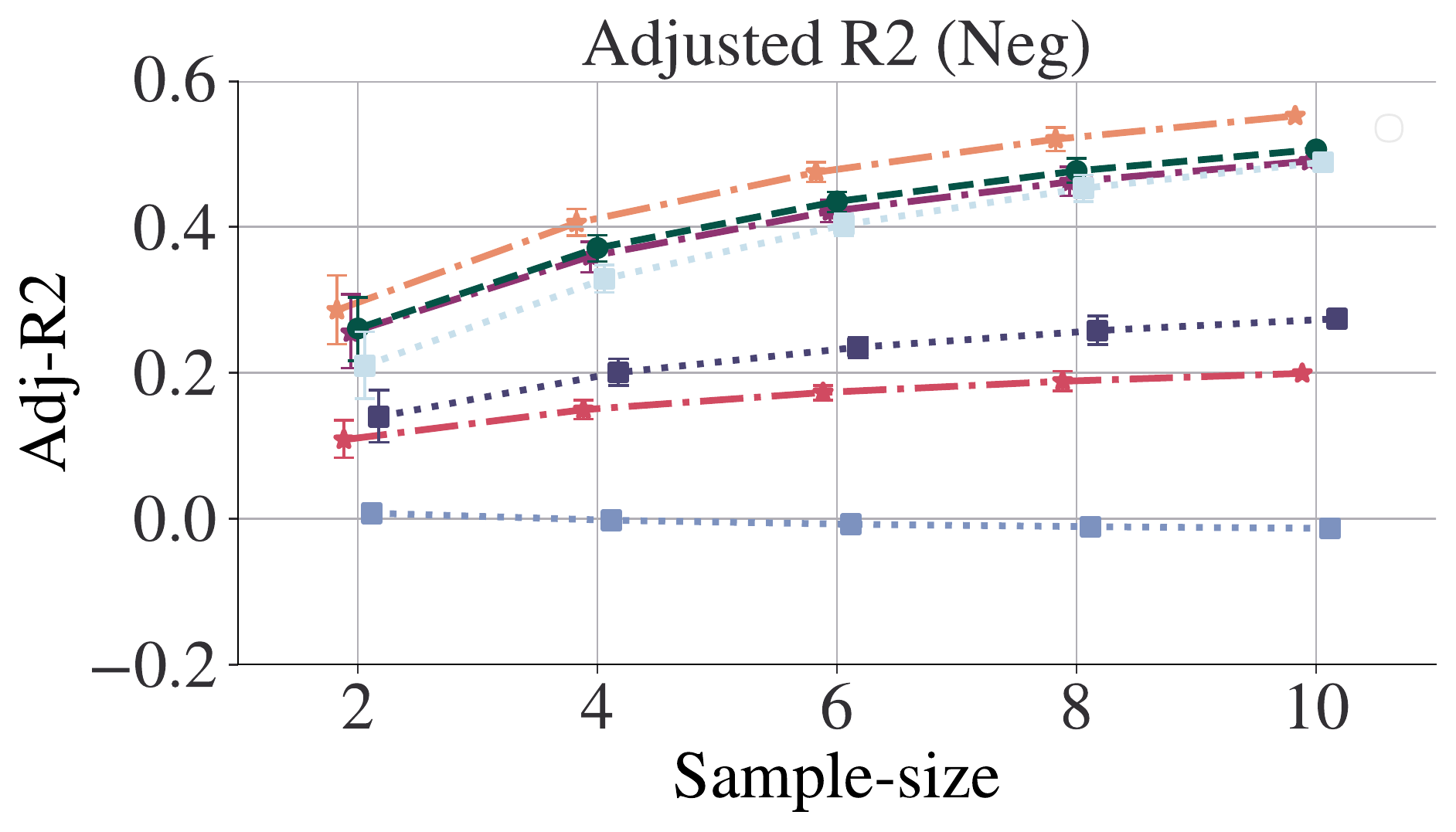}
    \caption{Influence of the number of clips on the correspondence of average CMOS to the objective measures.}
    \label{fig:size}
\end{figure}
\noindent
{\bf Crowd-size limits: }
Due to the high variance in the CMOS response and limits on the resources to expand the crowd-size, we analyse the CMOS responses on condition level as per Eq.{\ref{eq:CMOS_condition}}. While previous research has addressed the optimal number of responses per condition~\cite{naderi2020impact}, there is limited research on the influence of the number of clips per condition. We investigate how the predictability of the CMOS changes as the clip sample-size is increased~(2, 4, 6, 8, 10) while maintaining an equal number of clips by each speaker within the sample. The resulting mean coefficients and the standard deviation,  and the adjusted-$R^2$ over the combinations for each sample-size are illustrated in Fig.~\ref{fig:size}. We observe that \begin{enumerate*}\item The rate of change in the coefficients and adjusted-$R^2$ is similar for all objective measures. \item For the negative mode, sample size increase seems to have no effect on 3QUEST-N.   \item Since the rate-of-increase in coefficients and adjusted-$R^2$ is decreasing but not saturating as sample-size $\rightarrow 10$, we conclude $\#$clips $>10$ are necessary to obtain CCR-CMOS that correspond better to the state-of-the-art evaluation metric.
\end{enumerate*}

\section{Conclusions}
%\lipsum[1]
The goals of this work were to investigate the use of CCR to evaluate speech enhancement models and explore the parameters that are critical in designing such experiments. The findings are as follows: \begin{enumerate*}\item The bimodal distribution and the statistical correspondence between the positive and negative CMOS scores and the objective measures~(Fig.~\ref{fig:pos_neg_workers},~\ref{fig:distribution},  Tab.~\ref{tab:allPosNeg}, Tab.~\ref{tab:correlations}) indicate the need to analyse the positive-negative modes of responses separately or incorporate the positive-negative affiliation of the responses within the analysis.
The higher scale available in the CCR methodology allows to differentiate the scores into different groups based on the user's preference over noise reduction or quality of speech. \item A higher and significant correlation is calculated for positive workers and noise type, while the correlation is better between negative workers and the speech level~(Tab.~\ref{tab:effectofConditions}). SNR showed a significant effect over both positive and negative groups of CMOS responses. \item Condition-level analysis~(Fig.~\ref{fig:obj_vs_cmos}) showed higher statistical correspondence and hence predictability to the objective measures considered in this paper, in comparison to the clip/signal level analysis~(Tab.~\ref{tab:sta_clip_level}).
\end{enumerate*}

\begin{comment}
For future work, further testing of the CCR methodology must be researched by asking to rate specific features rather than overall quality.

\end{comment}
% higher discriminating scale

\balance
\bibliographystyle{IEEEbib}
%\nocite{*}

\bibliography{ref}

\end{document}